\documentclass[a4paper,fleqn]{cas-dc}

\usepackage[authoryear,longnamesfirst]{natbib}
\usepackage{cite}
\usepackage{amsmath,amssymb,amsfonts}
\usepackage{algorithmic}
\usepackage{graphicx}
\usepackage{textcomp}

\usepackage[]{xcolor}

\usepackage{balance}

\usepackage{pifont}
\usepackage{array}
\usepackage{booktabs}

\usepackage{hyperref}
\usepackage[capitalize,noabbrev]{cleveref}

\usepackage{amsmath}
\usepackage{enumitem}
\usepackage{multirow}

\usepackage{url}
\usepackage{etoolbox} 
\usepackage{tablefootnote}
\usepackage{listings}
\usepackage{todonotes}
\usepackage{verbatim}

\def\tsc#1{\csdef{#1}{\textsc{\lowercase{#1}}\xspace}}
\tsc{WGM}
\tsc{QE}
\tsc{EP}
\tsc{PMS}
\tsc{BEC}
\tsc{DE}

\renewcommand{\cite}{\citep}
\newcommand{\mycheck}{\ding{51}}%

\newcommand{\saveFeature}[3]{%
  \expandafter\newcommand\csname #1ID\endcsname{#1}%
  \expandafter\newcommand\csname #1FeatureID\endcsname{#2}%
  \expandafter\newcommand\csname #1Text\endcsname{#3}%
}

\newcommand{\featureText}[1]{%
    \ifcsname #1Text\endcsname
    \csname #1Text\endcsname
  \else
    \PackageError{FeatureManagement}{Text `#1' not defined}{}%
  \fi
}

\newcommand{\featureID}[1]{%
  \ifcsname #1FeatureID\endcsname
    \csname #1FeatureID\endcsname.%
  \else
    \PackageError{FeatureManagement}{FeatureID `#1' not defined}{}%
  \fi
}

\newcommand{\featureIDWithoutDotItalic}[1]{%
  \ifcsname #1FeatureID\endcsname
    \textit{\csname #1FeatureID\endcsname}%
  \else
    \PackageError{FeatureManagement}{FeatureID `#1' not defined}{}%
  \fi
}

\newcommand{\featureIDWithoutDot}[1]{%
  \ifcsname #1FeatureID\endcsname
    \csname #1FeatureID\endcsname%
  \else
    \PackageError{FeatureManagement}{FeatureID `#1' not defined}{}%
  \fi
}

\newcommand{\featureIDBold}[1]{%
  \textbf{\featureID{#1}}%
}

\newcommand{\featureIDUnderlined}[1]{%
  \underline{\featureID{#1}}%
}

\newcommand{\featureIDItalic}[1]{%
  \textit{\featureID{#1}}%
}

\newcommand{\sizea}{0.4cm}

\def\BibTeX{{\rm B\kern-.05em{\sc i\kern-.025em b}\kern-.08em
    T\kern-.1667em\lower.7ex\hbox{E}\kern-.125emX}}

\begin{document}
\let\WriteBookmarks\relax
\def\floatpagepagefraction{1}
\def\textpagefraction{.001}
\shorttitle{Mind the Gap: The Missing Features of the Tools to Support User Studies in Software Engineering}
\shortauthors{Costa, Barbosa, Cunha}

\title [mode = title]{Mind the Gap: The Missing Features of the Tools to Support User Studies}



\author[1]{Lázaro Costa}[orcid=0000-0002-6317-8792]
\cormark[1]
\fnmark[1]
\ead{lazaro@fe.up.pt}


\affiliation[1]{organization={HASLab/INESC TEC, Faculty of Engineering, University of Porto},                
                country={Portugal}}

\author[2]{Susana Barbosa}[orcid=0000-0003-2198-3715]
\ead{susana.a.barbosa@inesctec.pt}


\affiliation[2]{organization={INESC TEC},
                country={Portugal}}

\author[1]{Jácome Cunha}[orcid=0000-0002-4713-3834]
\ead{jacome@fe.up.pt}


\cortext[cor1]{Corresponding author}

\fntext[fn1]{The first author was supported by the doctoral Grant SFRH/BD/151366/2021 financed by the Portuguese Foundation for Science and Technology (FCT), and with funds from Portugal 2020, under MIT Portugal Program.}



\begin{abstract}
User studies are paramount for advancing science. However, researchers face several barriers when performing them despite the existence of supporting tools.
In this work, we study how existing tools and their features cope with previously identified barriers. Moreover, we propose new features for the barriers that lack support. We validated our proposal with 102 researchers, achieving statistically significant positive support for all but one feature.
We study the current gap between tools and barriers, using features as the bridge. We show there is a significant lack of support for several barriers, as some have no single tool to support them. 
\end{abstract}


\begin{highlights}
\item Current tools cannot fully aid with user studies
\item Several barriers faced by researchers have no tool support
\item We propose features to cope with all barriers
\item Researchers validated the proposed features
\end{highlights}


\maketitle

\section{Introduction} 
\label{sec:Userintro}

For many years, the research community has fostered the need for empirical evaluation of proposed techniques, methodologies, and corresponding tools. In particular, user studies are now standard when evaluating prototypes and implementing new approaches.
The design and execution of user studies and making all the collected information available are crucial to support research work and claims. Indeed, a significant amount of scientific research is currently supported by user studies. 
Moreover, empirical evaluation is also common in many software companies. While these studies are quite common, they are also difficult to design and run~\cite{Buse2011}.
Indeed, they require the execution of the study under specific (controlled) conditions and the capture of all the relevant information for further analysis.

However, it is not easy to plan, execute, capture, and make all the relevant information of a user study available~\cite{Freire2013}. Additionally, creating the required environment and centralizing all the collected information in a simple package for later analysis is challenging~\cite{Briand2004, Neto2015}. In previous work, \citet{Myers2023} presented a set of barriers and challenges reported by researchers when they were working on user studies.
They interviewed 26 researchers, who collectively reported 18 different barriers that they found during the diverse phases of user studies.




There are tools from the software engineering (SE) community available to aid researchers in their user studies, such as Ginger2~\cite{Ginger2}, SESE~\cite{Arisholm2002}, Experiment Manager~\cite{Hochstein2008}, eSEE~\cite{Travassos2008}, ARRESTT~\cite{Araujo2016}, ExpDSL~\cite{Haser2016, Haser2018}, or K-Alpha \cite{Silveira2012}.
However, these approaches have limitations in terms of recruiting participants, designing studies using wizards and guides, reusing previous studies, and reusing configurable task interfaces and components.
In the field of behavioral science, several tools (jsPsych~\cite{Leeuw2015}, LookIt~\cite{Scott2017} and Gorilla~\cite{Anwyl2020}) are able to solve these limitations, but they are specific to this domain.
We further discuss related work in \cref{sec:Userrelated}.

In previous work~\cite{vlhcc24}, we examined current tools, pinpointing specific features, which we summarize in \cref{sec:Usertools}. 
In this work, we seek to study whether such features can aid with the barriers previously identified by~\citet{Myers2023}. In particular, we answer the following research question (RQ):

\noindent
\textbf{RQ1: Are the current tools able to address all the barriers faced by researchers when conducting user studies?}

To answer this RQ, we bridge existing tools and barriers through the features we identified and validated, thus revealing the current gap~\cite{Myers2023} (\cref{sec:Userfeatures}).
Specifically, no tool can cope with all the barriers. In fact, there are barriers that no single tool addresses.

With this work, we also answer the following RQ:

\noindent
\textbf{RQ2: Which features can aid researchers in addressing the barriers faced when conducting user studies?}

To answer this RQ, we take the features of existing tools and match them with the previously identified barriers. Since the currently available features are not sufficient to overcome all barriers, we also propose new features.
We detail the relationships between barriers and features in \cref{sec:Userfeatures}.
We validated this match with 102 experienced researchers, who mainly agreed with our proposal. Indeed, all but one feature were statistically validated by the participants, as explained in \cref{sec:Uservalidation}.

On the basis of the answers to the RQs (see \cref{sec:Userdiscussion} for a discussion), our main contributions are as follows:
\begin{itemize}
    \item An overview of the state-of-the-art of the current tools and their features to support user studies.
    \item A list of community-validated features for addressing the barriers previously identified and the corresponding relationships between features and barriers.
    \item The identification of the existing gap between the current tools and the barriers faced by researchers.
\end{itemize}

In the next section (\ref{sec:Userrelated}), we review related work, and in \cref{sec:Userconclusion}, we present conclusions and future work.

%
%
%
%
%
%
%

\section{Related Work} 
\label{sec:Userrelated}


Our work builds on the work of 
\citet{Myers2023}, who presented a study in which 26 researchers were interviewed to understand their challenges when conducting programmer user studies. The study identifies several challenges, providing high-level abstract solutions to address them. 
On the other hand, our work focuses on the concrete tool features that software should have to reduce/resolve those barriers in the development of user studies. 

\citet{Buse2011} explore both the advantages and obstacles associated with user evaluation in SE research. Their comprehensive analysis sheds light on the benefits of user evaluation methodologies while addressing the challenges and limitations encountered in their implementation. This work provides valuable guidance for researchers seeking to optimize user evaluation practices in SE studies. However, it does not define a set of concrete characteristics that the software must contain to optimize user evaluation practices as we do in this work.

\citet{Sjoeberg2005} conducted a survey of controlled experiments in SE, offering a thorough examination of methodologies and outcomes in the field. However, this study provides insights into trends, challenges, and best practices surrounding controlled experiments in SE research while we propose concrete features that the tools should include. 

\citet{Kitchenham2002} propose a set of research guidelines intended to encourage discussion and collaboration among software researchers. Their primary objective is to offer valuable assistance to researchers, reviewers, and meta-analysts as they engage in various stages of empirical study design, execution, and evaluation within the field of SE. However, research guidelines are proposed to encourage debate and collaboration between researchers at various stages of empirical studies. They do not propose functionalities that the software must have to follow the proposed guidelines.


\citet{namoun2021review} conducted a comprehensive review, analyzing the research issues and challenges in automated website usability evaluation tools. They provide a detailed exploration of existing tools and introduce a usability framework comprising 19 usability dimensions, examining how 10 popular web usability testing tools align with this framework. However, this work focused on a specific domain (websites).

\citet{MARTIN2014302} offer guidance for the development of accessibility evaluation tools within the context of the Unified Software Development Process. It outlines a systematic approach to create tools that assess software accessibility. The guidance provided can aid developers in ensuring that accessibility considerations are seamlessly integrated into the software development life cycle. However, this is very focused work on a specific aspect (accessibility).

\citet{Ivie2018} delve into the technical challenges hindering reproducibility and replicability, review existing approaches, and highlight areas for further research. While one cannot reproduce an empirical study, it is possible to replicate it. Reproducing a user study would require running the exact same study under the exact same conditions, which is impossible as participants would have already done the study and thus would not be in the same state. However, we can discuss the replicability of an empirical study, meaning that one can execute it with different participants and achieve the same results. To do so, the software used, if any, would need to execute in the exact same way, which is one of the challenges faced by researchers as previously identified by \citet{Myers2023}. Thus, there is a clear connection between replicability and empirical studies. While \citet{Ivie2018} focus on reproducibility and replicability, our work focuses on user studies, which also often suffer from the same challenges. Thus, solutions for reproducibility and replicability may also solve some of the challenges of user studies.


\section{Background} 
\label{sec:Usertools}

\saveFeature{public} 
    {Public sharing} 
    {Public sharing user studies (including tasks, code, and results) to inspire others.}

\saveFeature{taskTemplates}
    {Task templates} 
    {Have pre-defined templates for task types to facilitate the creation of new tasks.}

\saveFeature{userStudiesTemplates}
    {User studies templates}
    {Pre-defined templates for types of user studies to facilitate the creation of new studies.}

\saveFeature{reuseAdapt}
    {Reuse and adaptation of user studies}
    {Reuse and adapt publicly available user studies to reduce the time spent developing a new study.}

\saveFeature{recordEnvironment}
    {Recording the user's environment}
    {Automatically record the user's environment (sound, screen, etc.) to allow future analysis by the researcher.}

\saveFeature{sendCollectedData}
    {Send the collected data}
    {Automatically send the collected data to the researcher to facilitate the data collection.}

\saveFeature{remoteExperiments}
    {Remote experiments}
    {Enable remote monitoring of users' studies in order to remotely track participants' progress and reduce face-to-face interaction with participants.}
    

\saveFeature{register}
    {Participant registration}
    {The platform allows participants to register on the platform to facilitate interaction between researchers and potential participants. It works online and contains several user studies accessible to the platform's community.}
    

\saveFeature{communication}
    {Researcher-participant communication}
    {Allow communication between researchers and participants for easier contact.}
    

\saveFeature{participantsPreferences}
    {Participants' preferences}
    {Set participants' preferences to only receive notifications from the platform about the opening of user studies considered relevant.}

\saveFeature{notifyParticipants}
    {Participants notification}
    {Notify registered participants of the opening of new user studies according to the proposed preferences of each participant.}
    

\saveFeature{package}  
    {Packaging of user studies}
    {All the necessary information from the study is packaged (code, data, etc.) so that they can be reproduced in the future.}

\saveFeature{privacyLevel}
    {Data privacy level}
    {Choose the privacy level for the data collected.}
    

\saveFeature{RecoverExperimentalData}
    {Recovery of experimental data}
    {Automatically recover the experiment data after a failure without data loss.}

\saveFeature{BackupExperimentalData}
    {Data backup}
    {Automatically backup the experimental data on a central server to prevent data loss.}

Our study is focused on a set of platforms that were identified as relevant by prior research on user studies for Software Engineering~\cite{Myers2023}.
In previous work, we identified a set of features currently available in tools that support user studies~\cite{vlhcc24}. 
Although we do not intend to replicate such work here, for completeness, we include the list of tools (\cref{subsec:tools}) and corresponding features (\cref{subsec:features}), as well as a brief description of each feature.

\subsection{Tools}\label{subsec:tools}

\paragraph*{Ginger2}  
This is an integrated environment designed to support \textit{in vitro} studies in empirical SE~\cite{Ginger2}. Combining hardware and software components enables the automatic collection of diverse types of data, including three-dimensional participant movements, skin resistance levels, mouse and keystroke activity, eye-tracking data, and video recordings. However, it is important to note that this platform was developed more than 25 years ago.
%

\paragraph*{SESE}  
Simula Experiment Support Environment (SESE) is a web-based platform designed to facilitate large-scale remote programmer user studies as well as face-to-face experiments~\cite{Arisholm2002}. It allows users to register on the platform, input their personal information, and recover experimental data in the event of a failure, ensuring no data are lost. Researchers can invite participants to study and use a built-in messaging system to interact with users (e.g., providing instructions). Additionally, SESE supports the management and monitoring of the collected data, with all the data backed up to a central server.
%

\paragraph*{Experiment Manager}  
This framework is designed to support SE experiments in high-performance computing (HPC) classroom settings~\cite{Hochstein2008}. Participants complete programming tasks under supervision, using either their local tools or integrated development environments (IDEs). Researchers can monitor the experimental progress and review the data collected during the study. At the conclusion of the experiment, the collected data are packaged and sent to a central server for analysis. Additionally, the framework facilitates data analysis while ensuring the secure and permanent deletion of sensitive data from datasets provided by external researchers.
%

\paragraph*{eSEE}  
The experimental Software Engineering Environment (eSEE) is a framework designed to support researchers throughout the entire SE experimentation process. This encompasses activities such as defining, planning, executing, and packaging studies within the field~\cite{Travassos2008}. Over time, however, maintaining the system became increasingly time-intensive and eventually unsustainable. To address this, a new platform was developed to preserve the data and packages~\cite{Santos2017}.

\paragraph*{ARRESTT}  
The Application of Reproducible Research on Evaluation of Software Testing Techniques (ARRESTT) framework provides comprehensive support for conducting, managing, and reproducing experimental studies focused on software testing techniques~\cite{Araujo2016}. By emphasizing reproducibility, ARRESTT enables researchers to systematically execute experiments and validate findings, fostering the advancement of reliable and repeatable research in the field of software testing.
%
%

\paragraph*{ExpDSL}  
This is an integrated, end-to-end tool designed to facilitate controlled human experiments via prototype domain-specific languages~\cite{Haser2016, Haser2018}. Built on the Meta Programming System (MPS), this tool provides comprehensive support for all phases of experimentation, offering an environment for designing, conducting, and managing experiments.

%

\paragraph*{K-Alpha}  
This is an extensible platform designed to support the management of SE experiments by addressing gaps in functionality through the use of plugins~\cite{Silveira2012}. Its primary goal is to facilitate the creation of reproducible experimental resources, enabling researchers to design, share, and replicate studies within the field of software testing. This includes support for techniques such as test case selection and test suite minimization.

\paragraph*{jsPsych}  
This is a JavaScript library designed for creating and managing online behavioral experiments by structuring them as a series of sequential steps and decisions~\cite{Leeuw2015}. Each step provides a customizable task interface, allowing participants to perform specific activities, with the results influencing subsequent stages of the experiment. The library includes a core set of task interfaces tailored for behavioral science research, which are openly available to the research community.

\paragraph*{LookIt}  
This is a collaborative online platform for remote data collection that enables researchers from different organizations to design and conduct experiments~\cite{Scott2017}. It allows researchers to outline the sequential steps and decisions within an experiment, providing a variety of customizable task interfaces for participants. Additionally, participants can join a shared pool, giving them the flexibility to engage in upcoming experiments at their preferred frequency.

\paragraph*{Gorilla}  
This is an online experiment builder that allows multiple researchers to design and host experiments~\cite{Anwyl2020}. Using its graphical user interface, researchers can configure task interfaces for their studies. Additionally, Gorilla provides a set of resources and guidance materials to help researchers create well-structured and effective experiments.

\subsection{Features}\label{subsec:features}

In \cref{tab:experimentsTools}, we consolidate and present a comprehensive overview of the features available across various tools evaluated in our study. Each row in the table lists a distinct feature along with the tools that support it. For completeness, we briefly describe each feature:

\begin{description}
\item[\featureID{public}] Facilitate the public sharing of user studies, including tasks, code, and results, to inspire and support other researchers.
\item[\featureID{taskTemplates}] Provide pre-defined templates for various task types, aiding in the creation of new experimental tasks.
\item[\featureID{userStudiesTemplates}] Offer pre-defined templates for different types of user studies to simplify the creation of new studies.
\item[\featureID{reuseAdapt}] The reuse and adaptation of publicly available user studies can be enabled to save time and effort when new experiments are designed.
\item[\featureID{recordEnvironment}] Automatically record aspects of the user’s environment, such as sound and screen activity, to enable detailed future analysis by researchers.
\item[\featureID{sendCollectedData}] The collected data are automatically transmitted to researchers, facilitating data collection.
\item[\featureID{remoteExperiments}] Support the remote monitoring of user studies, allowing researchers to track participants' progress without requiring face-to-face interactions.

\item[\featureID{register}] Allow participants to register on the platform, providing access to available user studies and facilitating interaction between researchers and potential participants.

\item[\featureID{communication}] By enabling direct communication between researchers and participants, easier interactions and clarifications can be obtained.

\item[\featureID{participantsPreferences}] The participants are allowed to set preferences, enabling them to receive notifications about user studies that align with their interests.

\item[\featureID{notifyParticipants}] The registered participants were notified about the availability of new user studies tailored to their selected preferences.

\item[\featureID{package}] All necessary study information, including code, data, and materials, are compiled into a package to ensure replicability and future reference.

\item[\featureID{privacyLevel}] Customizable privacy settings are offered for the collected data to align with ethical standards and participant consent.

\item[\featureID{RecoverExperimentalData}] Automatically recover experimental data following a failure, ensuring that no data are lost.

\item[\featureID{BackupExperimentalData}] An automatic backup of experimental data on a secure central server is provided to prevent data loss and ensure data integrity.
\end{description}

During this evaluation, we identified a total of 15 different features. 
As shown in \cref{tab:experimentsTools}, some features are available in several tools; for instance, 
\featureIDWithoutDotItalic{remoteExperiments} are offered by nine tools. 
On the other hand, \featureIDWithoutDotItalic{communication}, \featureIDWithoutDotItalic{privacyLevel} and \featureIDWithoutDotItalic{RecoverExperimentalData} 
are offered only by one tool each.
The web-based SESE platform is a tool that covers more features -- 8 of the 15.
As demonstrated, no current tool covers all the features.

In the next section, we discuss to what extent the current features are able to address the barriers encountered by researchers \cite{Myers2023}. 


\begin{table}[htbp]
\rowcolors{2}{gray!20}{white}
  \centering
  \caption{Features and list of tools implementing them}
    \begin{tabular}{p{3.5cm}p{3.9cm}}
    \rowcolor{gray!50}
    \textbf{Feature} & \textbf{Tools} \\
    \featureID{public} & ARRESTT \\ 
    \featureID{taskTemplates} & jsPsych \\ 
    \featureID{userStudiesTemplates} & Gorilla \\
    \featureID{reuseAdapt} & eSEE; ARRESTT \\
    \featureID{recordEnvironment} & LookIt \\
    \featureID{sendCollectedData} & SESE; Experiment Manager \\
    \featureID{remoteExperiments} & SESE; Experiment Manager; eSEE; ARRESTT; ExpDSL; K-Alpha; jsPsych; LookIt; Gorilla \\
    \featureID{register} & SESE; Experiment Manager; K-Alpha; jsPsych; LookIt; Gorilla \\
    \featureID{communication} & SESE \\
    \featureID{participantsPreferences} & SESE; K-Alpha; LookIt \\
    \featureID{notifyParticipants} & SESE; K-Alpha; LookIt \\
    \featureID{package} & Experiment Manager; eSEE; ARRESTT; ExpDSL; K-Alpha; jsPsych; Gorilla \\
    \featureID{privacyLevel} & LookIt \\
    \featureID{RecoverExperimentalData} & SESE \\
    \featureID{BackupExperimentalData} & SESE; jsPsych \\
    \bottomrule
    \end{tabular}%
  \label{tab:experimentsTools}%
\end{table}%

\section{Barriers Meet Features} 
\label{sec:Userfeatures}
\saveFeature{userStudiesFeedback-Our}
    {User studies feedback}
    {Enable the submission of user studies within the platform to enable feedback from other researchers before the study is implemented.}

\saveFeature{piloting-Our}
    {Piloting alternative designs}
    {Possibility of piloting alternative designs to help choose the most suitable one.}
    
\saveFeature{IRBTemplates-Our}
    {IRB templates}
    {Have pre-defined templates for IRB requirements to assist the process.}
    
\saveFeature{AutomaticIRBInformation-Our}
    {Extract user study data for the IRB}
    {Automatic extraction, from the user study, of information to be used in the IRB process.}
    
\saveFeature{promptReExecution-Our}
    {Enabling prompt re-execution}
    {Enabling prompt re-execution of the user study to allow quick and incremental evaluation of tools.}
    
\saveFeature{offerGuides-Our}
    {Offer guides for designing}
    {Offer guides for designing user studies, e.g., tutorials for easily constructing a new study.}
    
\saveFeature{PrePlanesUserStudy-Our}
    {Execution of the pre-planned use study}
    {Allow automatic execution of a pre-planned use study to reduce manual interventions.}
    
\saveFeature{containerization-Our}
    {Containerization of user studies}
    {Easy containerization of the user study. Containers are portable and isolated applications, facilitating migration between different infrastructures.}
    
As highlighted in the introduction, the previous study by \citet{Myers2023}, which involved interviews with 26 researchers, identified significant barriers in conducting programmer user studies. 
They have also proposed high-level possible solutions to reduce the 18 barriers found.

In our work, we propose concrete features that tools should implement to address these barriers. This can be seen as a requirement engineering exercise that others can use to improve their current tools or build new tools.
While there may be other interesting or useful features, our goal was not to brainstorm arbitrarily but to propose features that specifically address barriers previously identified in the literature regarding the definition and execution of user studies.

We start by describing each barrier (\textit{$B_i$}) and present a set of features to aid researchers in overcoming it, as well as the rationale that supports the pairing  (\cref{sec:Usernew_features}). In some cases, we group several barriers together because the corresponding features are the same.
However, as we shall see (\cref{sec:Usernew_features}), the available tools and features are insufficient to cope with all the identified barriers. Thus, we reuse the existing features presented in \cref{sec:Usertools} and introduce new features to address all the identified barriers. In this section, we attribute specific markings—existing features are in italics, and new features are underlined—to help the reader distinguish them.

We then present an empirical study in which we validated our proposal with 102 researchers through an online survey (\cref{sec:Uservalidation}). 

\subsection{Detailed Analysis of Barrier-Feature Interactions}\label{sec:Usernew_features}
We now present each barrier, denoted by $B_i$, as originally introduced by \citet{Myers2023} and the features that we propose to cope with them, explaining why each feature addresses the barrier.

    
\noindent
\textit{$B_{1}$:} Task design is hard. Task design frequently requires devising a solution that meets various requirements. Crafting a task that fulfills all these requirements can be especially challenging. 
        \begin{description}
                
                \item[\featureIDItalic{public}] Publicly sharing task designs, including methodologies, requirements, and outcomes, provides researchers with access to proven examples and diverse approaches to similar challenges. This transmission fosters collaboration and knowledge exchange within the community, enabling researchers to learn from successful designs and avoid common pitfalls. By leveraging shared insights, researchers can more effectively address complex requirements in task design, improving both efficiency and efficacy.
                
                \item[\featureIDItalic{taskTemplates}] Offering task templates provides a structured and reliable way to design tasks that meet diverse and often complex requirements. Templates highlight essential elements such as objectives, constraints, and evaluation criteria, guiding researchers through the design process. By reducing the cognitive load associated with starting from scratch, templates allow researchers to focus on refining their tasks and save significant time.
                
                \item[\featureIDUnderlined{userStudiesFeedback-Our}] 
                Feedback from other researchers is essential for refining task designs to meet specific requirements and ensure their clarity and effectiveness. Indeed, researchers often do this in an \textit{ad hoc} way with colleagues. Constructive input helps identify potential weaknesses, ambiguous elements, or gaps in the design, allowing researchers to make improvements. By incorporating diverse perspectives, researchers can enhance the quality, relevance, and robustness of their tasks, leading to better experimental outcomes and more reliable results.
        \end{description}

\noindent
\textit{$B_{2}$:} It is difficult to understand studies' design trade-offs in advance. Finding the right balance for the research question is not always easy. It requires creating realistic and limited tasks so that the effect under study is accurately measured and not confounded by other variables.
        \begin{description}
                
                \item[\featureIDUnderlined{userStudiesFeedback-Our}] The incorporation of feedback early in the task design process provides researchers with valuable insights into the effectiveness of various design elements. Feedback helps highlight potential trade-offs and allows researchers to evaluate how well specific aspects of the design align with their research objectives. By understanding the impact of these trade-offs early, researchers can refine tasks to minimize confounding variables and better target the desired effects, thus improving the reliability of the study's results.
               
                \item[\featureIDItalic{userStudiesTemplates}] Templates for user studies provide a structured starting point, guiding researchers through the process of designing the study.
                One can imagine a set of typical user studies' designs (e.g., between-subject, within-subject) the researcher can start with, thus ensuring that the design is correct (e.g., if one chooses a within-subject, then the order of the treatments should be randomized).
                
                \item[\featureIDUnderlined{piloting-Our}] 
                Piloting alternative task designs allows researchers to explore and evaluate different trade-offs before committing to a final design. Testing multiple variations helps uncover which designs effectively address the research question while avoiding confounding variables. This iterative process provides critical insights into the strengths and weaknesses of each approach, enabling researchers to make informed decisions and achieve the optimal balance for their study, ultimately enhancing the validity of their findings.
        \end{description}

\noindent
\textit{$B_{3}$:} Researchers cannot take studies off the shelf for reuse. It is a challenge to reuse/adapt a previous user study because of the lack of information on how the user study was carried out, making it necessary to repeat the entire user study process again.

        \begin{description}
                    \item[\featureIDItalic{public}] Publicly sharing user studies' designs enhances transparency and facilitates knowledge exchange within the research community. By making detailed study designs, execution steps, and results accessible, researchers have enabled others to understand and replicate successful approaches. This openness reduces the need to start from scratch, empowering researchers to adapt existing studies more efficiently and build on prior work.
                    
                    \item[\featureIDItalic{reuseAdapt}] Establishing strategies for reusing and adapting user studies enhances research efficiency. By modifying elements of existing studies (designs, tasks, or protocols), researchers can address new research questions without duplicating effort. This approach saves time and resources, enabling researchers to focus on refining their studies.
                    
                    \item[\featureIDItalic{userStudiesTemplates}] Standardized and customizable templates for user studies provide a structured framework that simplifies the design process. By outlining critical components, such as research objectives, participant criteria, and procedures, templates help researchers create studies that are both ground-based and reproducible. 
                    
                    \item[\featureIDItalic{taskTemplates}] Task templates provide pre-defined structures for designing tasks, including detailed instructions, evaluation criteria, and layout guidelines. By promoting best practices and reducing the effort required to design tasks from scratch, templates enable researchers to focus on addressing specific research objectives.
        \end{description}

\noindent
\textit{$B_{4}$:} Institutional Review Board (IRB) requires quite some effort to approve studies.
        \begin{description}
                \item[\featureIDUnderlined{IRBTemplates-Our}]
                 Providing IRB templates helps reduce the complexity and effort required to prepare proposals for ethical review. These templates outline the essential components needed by IRBs, such as study objectives, participant recruitment strategies, informed consent documents, and data management plans. By offering a clear structure, templates enable researchers to efficiently compile the necessary information, ensuring compliance with ethical standards while minimizing the time and effort spent on the approval process.
                 
                \item[\featureIDUnderlined{AutomaticIRBInformation-Our}] 
                 Automating or structuring the extraction of relevant user study data simplifies the preparation of IRB submissions. Researchers often need to provide detailed documentation on participants, methodologies, and potential risks. By integrating a feature that extracts this information directly from study designs or protocols, researchers can quickly assemble accurate and comprehensive IRB submissions. This approach not only saves time but also enhances transparency and consistency in the ethical review process.
        \end{description}

\noindent
\textit{$B_{5}$:} Building and integrating tools is challenging. The effort and time required to set up the environment to evaluate a novel tool are excessively high. This barrier is specific to studies that include some software prototypes.
        \begin{description}
                \item[\featureIDUnderlined{promptReExecution-Our}] 
                Automating the deployment and evaluation process through pipelines reduces the time and effort required to set up environments for evaluating tools. By integrating tool evaluation into automated testing workflows, researchers can eliminate repetitive tasks, allowing faster iterations and adjustments to tool configurations. This automation provides immediate feedback on tool performance and its impact on user studies, allowing researchers to focus on refining their tools and achieving incremental improvements without excessive setup efforts.
        \end{description}

\noindent
\textit{$B_{6}$:} It is difficult to collect data correctly. 
The execution time of the tasks is often easy to obtain. However, 
some measures involve more challenging logistical processes (e.g., recording the screen, saving the video, and associating it with the participant for further analysis).
        \begin{description}
                \item[\featureIDItalic{recordEnvironment}] Enabling automated recording of the user’s environment during task execution (e.g., capturing screen activity, logging user interactions, recording system metrics in real-time) facilitates efficient data collection. This feature eliminates the need for manual tracking of complex data points, allowing researchers to gather comprehensive, synchronized data on task performance. By integrating these recordings into the study setup, researchers ensure that valuable information is captured consistently and is readily available for analysis.
                
                \item[\featureIDItalic{sendCollectedData}] Implementing automated data transfer mechanisms ensures that collected data, such as screen recordings and log files, are sent to designated storage or analysis platforms. This reduces the manual effort involved in handling and organizing data while also ensuring that the data are transferred securely and remain accessible for further examination. The automation of this process guarantees that data integrity is maintained and allows researchers to focus on the analysis phase.
        \end{description}

\noindent
\textit{$B_{7}$:} Researchers often feel that they are not prepared to design and run a user study because they lack knowledge (especially for new researchers). \textit{$B_{8}$:} Gaining the needed knowledge is inefficient. 
        \begin{description}
                \item[\featureIDUnderlined{userStudiesFeedback-Our}] Establishing a feedback loop with experienced practitioners or researchers enables less-experienced researchers to gain valuable insights into study design, execution, and result interpretation. This ongoing exchange of knowledge helps bridge gaps in understanding, enabling researchers to learn best practices in empirical evaluation. By leveraging the expertise of others, researchers can increase their proficiency, reducing the challenges posed by a lack of knowledge.
                
                \item[\featureIDItalic{userStudiesTemplates}] Providing standardized templates for designing user studies offers a practical framework for researchers, particularly those unfamiliar with empirical evaluation. These templates highlight essential components, methodologies, and best practices, serving as a step-by-step guide for creating well-structured studies. By using these templates, researchers can quickly familiarize themselves with the process and avoid common pitfalls, making the acquisition of foundational knowledge more accessible and efficient.
                
                \item[\featureIDUnderlined{offerGuides-Our}]
                 Developing detailed guides or tutorials for empirical evaluation provides researchers with a structured pathway for acquiring critical knowledge. Tailoring these resources to different proficiency levels ensures accessibility for beginners while also offering advanced techniques for experienced researchers. Comprehensive guides help researchers build confidence, improve their methodological rigor, and overcome the inefficiencies often associated with learning empirical evaluation practices independently.
        \end{description}

\noindent
\textit{$B_{9}$:} Some researchers feel uncomfortable with people.
        \begin{description}
                \item[\featureIDItalic{remoteExperiments}] Allowing participants to engage in study activities remotely reduces the need for face-to-face interaction, making the process more comfortable for researchers who may feel uneasy in direct participant interactions. By enabling data collection and monitoring from a distance, remote experiments provide an effective way to conduct user studies while accommodating researchers’ preferences.
                
                \item[\featureIDItalic{register}] Implementing a participant registration system facilitates the recruitment and management of participants without requiring in-person interactions. Researchers can efficiently handle participant information, schedule sessions, and manage communication through the platform, minimizing discomfort associated with face-to-face recruitment and administrative tasks.
                
                \item[\featureIDItalic{communication}] Digital communication channels provide a professional and comfortable means for researchers to interact with participants. These channels allow researchers to deliver instructions, address questions, and gather feedback without needing direct, in-person contact. This approach ensures effective collaboration while respecting researchers’ preferences for limited face-to-face interaction.
        \end{description}

\noindent
\textit{$B_{10}$:} It is difficult to recruit enough representative participants. \textit{$B_{11}$:} It is difficult to manage participants over time. \textit{$B_{12}$:} The recruitment material norms vary by study/organization. \textit{$B_{13}$:} It is difficult to select participants with the desired characteristics.
        \begin{description}
                \item[\featureIDItalic{register}] Creating a registration system where individuals can sign up for research studies simplifies the process of recruiting participants. By collecting key details such as demographics, background, and availability, this system allows researchers to access a pool of potential participants and efficiently identify those who meet the requirements of their studies. This approach addresses the challenge of recruiting enough representative participants and selecting individuals with desired characteristics, aiding in the recruitment process.
                
                \item[\featureIDItalic{participantsPreferences}] Enabling participants to specify their preferences and characteristics provides researchers with detailed insights into their suitability for different studies. This feature makes it easier to match participants with studies that align with their profiles, ensuring greater relevance and engagement. By incorporating participant preferences, researchers can address the difficulty of selecting participants with desired traits while building a more organized and responsive recruitment process.
                
                \item[\featureIDItalic{notifyParticipants}] A notification system helps keep participants engaged by informing them about upcoming studies, updates, or opportunities tailored to their preferences. By reaching out to participants directly, researchers can maintain their interest and involvement over time, improving the likelihood of recruiting enough representative participants. This feature also supports better study management by ensuring timely communication and fostering stronger connections between researchers and participants.
        \end{description}

\noindent
\textit{$B_{14}$:} End-to-end orchestration is cumbersome. The logistical process of assigning participants to tasks/groups is hard.
        \begin{description}
            \item[\featureIDUnderlined{PrePlanesUserStudy-Our}]
            Implementing a system for pre-planning user studies allows researchers to define tasks, create study groups, and assign participants in advance. By handling parameters such as scheduling and participant criteria early, this feature simplifies the orchestration of complex studies. It reduces the logistical challenges associated with coordinating participants and tasks, enabling researchers to focus on analysis and outcomes rather than technical/administrative burdens.

        \end{description}

\noindent
\textit{$B_{15}$:} Prototype software is not ready for deployment. \textit{$B_{16}$:} Deploying to a participant’s local personal computer (PC) is challenging. \textit{$B_{17}$:} Deploying to hosted virtual machines (VMs) is challenging. \textit{$B_{18}$:} Deploying to the web is challenging.
        \begin{description}
            \item[\featureIDUnderlined{containerization-Our}]
            Leveraging containerization technology, such as Docker, allows researchers to package prototype software along with its dependencies into a standardized and portable environment. Containers simplify deployment by ensuring consistent functionality across different platforms, whether on participants’ local PCs, hosted VMs, or the web. This approach addresses challenges related to incomplete prototype readiness, compatibility issues, and complex setups, enabling researchers to deliver software to participants with minimal effort and reducing the risk of errors.
        \end{description}

\subsection{Relation Between Barriers and Tools}
We seek to understand the barriers addressed by each existing tool, and thus, we relate the barriers to the tools through the features identified in \cref{sec:Usertools}.
That is, for each barrier, we want to understand if it is supported
by a certain tool. Thus, we relate the barriers with the tools using the features as the connection between both.

In \cref{tab:barriersFeaturesTools}, we provide a detailed comparison between the identified barriers (rows) and the features developed to overcome them, as well as the existing tools that incorporate these features (columns). Although \cref{tab:experimentsTools} outlines the features of the current tools, it does not comprehensively cover the features designed to address all identified barriers.

Each cell in the table is marked if the corresponding tool integrates the associated feature, providing a clearer picture of the gaps in existing solutions and highlighting the features that should be prioritized for future tool development.


\begin{table*}[!htb]
\setlength\tabcolsep{2pt}
  \centering
  \caption{The features of each tool and which barriers each address}
    \begin{tabular}{m{1.1cm}>{\raggedright}m{3.5cm}>{\columncolor{gray!20}\centering}m{1.2cm}>{\columncolor{white}\centering}m{0.81cm}>{\columncolor{gray!20}\centering}m{1.61cm}>{\columncolor{white}\centering}m{0.851cm}>{\columncolor{gray!20}\centering}m{1.51cm}>{\centering}m{1.2cm}>{\columncolor{gray!20}\centering}m{0.85cm}>{\centering}m{1.1cm}>{\columncolor{gray!20}\centering}m{0.95cm}>{\centering\arraybackslash}m{0.9cm}}
    \toprule
    \textbf{Barriers} & \textbf{Features} & \textbf{Ginger2} & \textbf{SESE} & \textbf{Experiment Manager} & \textbf{eSEE}  & \textbf{ARRESTT} & \textbf{ExpDSL} & \textbf{K-Alpha} & \textbf{jsPsych} & \textbf{LookIt} & \textbf{Gorilla} \\
    \midrule
    \multirow{3}[0]{*}{$B_{1}$} & \featureID{public} &   &   &   &   & \mycheck &   &   &   &   &  \\
          & \featureID{taskTemplates} &   &   &   &   &   &   &   & \mycheck &   &  \\
          & \featureID{userStudiesFeedback-Our} &   &   &   &   &   &   &   &   &   &  \\
          \midrule
    \multirow{4}[0]{*}{$B_{2}$} & \featureID{userStudiesFeedback-Our} &   &   &   &   &   &   &   &   &   &  \\
          & \featureID{userStudiesTemplates} &   &   &   &   &   &   &   &   &   & \mycheck \\
          & \featureID{piloting-Our} &   &   &   &   &   &   &   &   &   &  \\
          \midrule
    \multirow{5}[0]{*}{$B_{3}$} & \featureID{public} &   &   &   &   & \mycheck &   &   &   &   &  \\
          & \featureID{reuseAdapt} &   &   &   & \mycheck    & \mycheck &   &   &   &   & \mycheck \\
          & \featureID{userStudiesTemplates} &   &   &   &   &   &   &   &   &   & \mycheck \\
          & \featureID{taskTemplates} &   &   &   &   &   &   &   &   
          \mycheck &   &  \\
          \midrule
    \multirow{2}[0]{*}{$B_{4}$} 
         & \featureID{IRBTemplates-Our} &   &   &   &   &   &   &   &   &   &  \\
          & \featureID{AutomaticIRBInformation-Our} &   &   &   &   &   &   &   &   &   &  \\
          \midrule
    $B_{5}$ & \featureID{promptReExecution-Our} &   &   &   &   &   &   &   &   &   &  \\
          \midrule
    \multirow{2}[0]{*}{\parbox{3.1cm}{$B_{6}$}} & \featureID{recordEnvironment} &   &   &   &   &   &   &   &   & \mycheck &  \\
          & \featureID{sendCollectedData} &   & \mycheck & \mycheck &   &   &   &   &   &   &  \\
          \midrule
          
    \multirow{4}[0]{*}{\parbox{3.1cm}{$B_{7}$\\ $B_{8}$}} & \featureID{userStudiesFeedback-Our} &   &   &   &   &   &   &   &   &   &  \\
          & \featureID{userStudiesTemplates} &   &   &   &   &   &   &   &   &   & \mycheck \\
          & \featureID{offerGuides-Our} &   &   &   &   &   &   &   &   &   &  \\
          \midrule
    \multirow{4}[0]{*}{\parbox{3.1cm}{$B_{9}$}} & \featureID{remoteExperiments} &   & \mycheck & \mycheck & \mycheck    & \mycheck & \mycheck    & \mycheck    & \mycheck & \mycheck & \mycheck \\
          & \featureID{register} &   & \mycheck & \mycheck &   &   &   & \mycheck    & \mycheck & \mycheck & \mycheck \\
          & \featureID{communication} &   & \mycheck &   &   &   &   &   &   &   &  \\
          \midrule
    \multirow{4}[0]{*}{\parbox{3.1cm}{$B_{10}$\\ $B_{11}$ \\ $B_{12}$ \\ $B_{13}$ }} & \featureID{register} &   & \mycheck & \mycheck &   &   &   & \mycheck    & \mycheck & \mycheck & \mycheck \\
          & \featureID{participantsPreferences} &   & \mycheck &   &   &   &   & \mycheck    &   & \mycheck &  \\
          & \featureID{notifyParticipants} &   & \mycheck &   &   &   &   & \mycheck    &   & \mycheck &  \\
& &   &   &   &   &   &   &   &   &   &  \\
          
          \midrule
    $B_{14}$ & \featureID{PrePlanesUserStudy-Our} &   &   &   &   &   &   &   &   &   &  \\

          \midrule

    \multirow{2}[0]{*}{\parbox{3.1cm}{\vspace{-.15cm}$B_{15}$ \\ $B_{16}$ \\ $B_{17}$ \\ $B_{18}$ }}  & \featureID{containerization-Our} &   &   &   &   &   &   &   &   &   &  \\
    & &   &   &   &   &   &   &   &   &   &  \\
    & &   &   &   &   &   &   &   &   &   &  \\
          
          \midrule
    -- & \featureID{package} &   &   & \mycheck & \mycheck    & \mycheck & \mycheck    & \mycheck    & \mycheck &   & \mycheck \\
      -- & \featureID{privacyLevel} &   &   &   &   &   &   &   &   & \mycheck &  \\
       --   & \featureID{RecoverExperimentalData} &   & \mycheck &   &   &   &   &   &   &   &  \\
     --   & \featureID{BackupExperimentalData} &   & \mycheck &   &   &   &   &   & \mycheck &   &  \\
     

    
          \bottomrule
    \end{tabular}%
  \label{tab:barriersFeaturesTools}%
\end{table*}%

Notably, the table is quite sparse, meaning that there is
little support from current tools for the barriers researchers encounter.
Among the 18 barriers, seven (approximately 40\%) do not have a single tool 
offering support for researchers. These barriers are $B_4$ (IRB's related issues), $B_5$ (tool development), $B_{14}$ (orchestration), and $B_{15}$ through $B_{18}$ (deployment).
For many of the other barriers, the scenario is not ideal. $B_2$ (trade-offs), $B_7$, and $B_8$ (both related to knowledge) have just one tool to address them, $B_1$ (task design) has two, $B_6$ (data collection) has three, and $B_3$ (studies out of the shelf) has four.
On the other hand, barriers $B_9$ (researcher-participant relationship) and $B_{10}$ through $B_{13}$ (all related to participant recruitment) are well covered with nine and six tools, respectively. 

From the tools’ perspective, Ginger2 does not offer any
support, ExpDSL tackles a single barrier, eSEE supports two, ARRESTT three, K-Alpha five,  SESE, Experiment Manager, and LookIt six, jsPsych eight, and Gorilla nine. This shows that tools from the field of behavioral science are the ones most prepared to cope with user studies.
Note that the support of each tool may be partial, depending on the tool's features.

\subsection{Answering RQ1}\label{sec:Userrq1}
On the basis of \cref{tab:barriersFeaturesTools} and our analysis, we can answer RQ1.
Existing tools can help researchers with approximately 60\% of the 18 identified barriers. However, no tool addresses all barriers -- the best tools address eight or nine barriers, at most, that is, only 50\%. 
Moreover, this support is, in many cases, partial, as the tools implement just a fraction of the features researchers find relevant to tackling the corresponding barriers.
Thus, there is indeed a gap between the current tools offered and the barriers researchers face. More work is necessary to devise a tool or improve existing tools to fully support researchers when performing their user studies.

In the next section, we present the results of our survey, which assesses whether researchers support the proposed features as viable solutions to overcome user studies' barriers.

\section{Evaluation}\label{sec:Uservalidation}
In this section, we present an empirical study to evaluate the features we propose to address the discussed barriers, as detailed 
in \cref{sec:Userfeatures}. The results provide empirical validation of these features and reveal that while most were widely supported, some were not. This distinction is crucial in understanding which features should be prioritized in future tool development.

\subsection{Study Design}

\paragraph{Hypotheses}
Our family of hypotheses is that, for each feature, participants consider that it helps researchers address the barriers that we associate with the feature. We further elaborate on the null hypotheses in \cref{sec:Userhyptest}.

\paragraph{Objects and Participants}
We carried out an online survey\footnote{Available at \url{https://forms.gle/AyTkB2X5FR2p72kJ6}.} implemented via Google Forms, in which the main objective was to validate with the research community the relevance of the proposed features to reduce the previously identified barriers in the development of user studies. 
We spread the survey on social media platforms (Twitter, Facebook, and LinkedIn) through public posts. Additionally, we distributed the survey via email to the authors of published works in the IEEE Symposium on Visual Languages and Human-Centric Computing (VL/HCC) 2023, in the IEEE/ACM International Conference on Software Engineering (ICSE) 2022, and in the ACM Conference on Human Factors in Computing Systems (CHI) 2023. We chose these venues because they represent major conferences where empirical studies are often presented, and thus, it is more likely that we would receive expert participants.  
No compensation was offered to the participants. 

\paragraph{Instrumentation}
We started by collecting demographic information about the participants and their research experience, represented by the number of scientific works published reporting a user study and their experience with user studies. Then, the participants were presented with a set of questions where each question introduced a barrier (or set of barriers when the features are the same) and the features we propose to address it, similar to what we have just done in \cref{sec:Userfeatures}.
Participants were asked to evaluate their level of agreement with each proposed feature using a five-point Likert scale~\cite{likert1932}. Importantly, they were instructed to assess whether each feature could help reduce or address the corresponding barrier, rather than expecting it to eliminate the issue completely. Additionally, the survey did not require participants to assume that all features were fully implemented from the outset. For example, a complete set of task templates is not necessary at launch; instead, the system is designed to allow researchers to contribute and expand templates over time. They also had the possibility of leaving comments and suggestions in free-text form so that they could express their thoughts about each individual proposal. This is possible for each individual barrier (or set).

\paragraph{Execution}
The survey was available online between February and March 2024. Most answers were obtained in the first few weeks. As time passed, we received fewer and fewer answers, so we decided to close the survey.

\paragraph{Data Collection}
We collected all the data via a single online form.

\paragraph{Analysis}
After data collection, we proceeded with descriptive statistics.
We also analyzed the free-form feedback provided by the participants.

\subsection{Results} \label{subsection:features-results}
We now present the results of our study in detail.

\subsubsection{Descriptive Statistics}

\subsubsection*{Participants}
\begin{figure}[!htb]
\centerline{\includegraphics[width=\columnwidth] {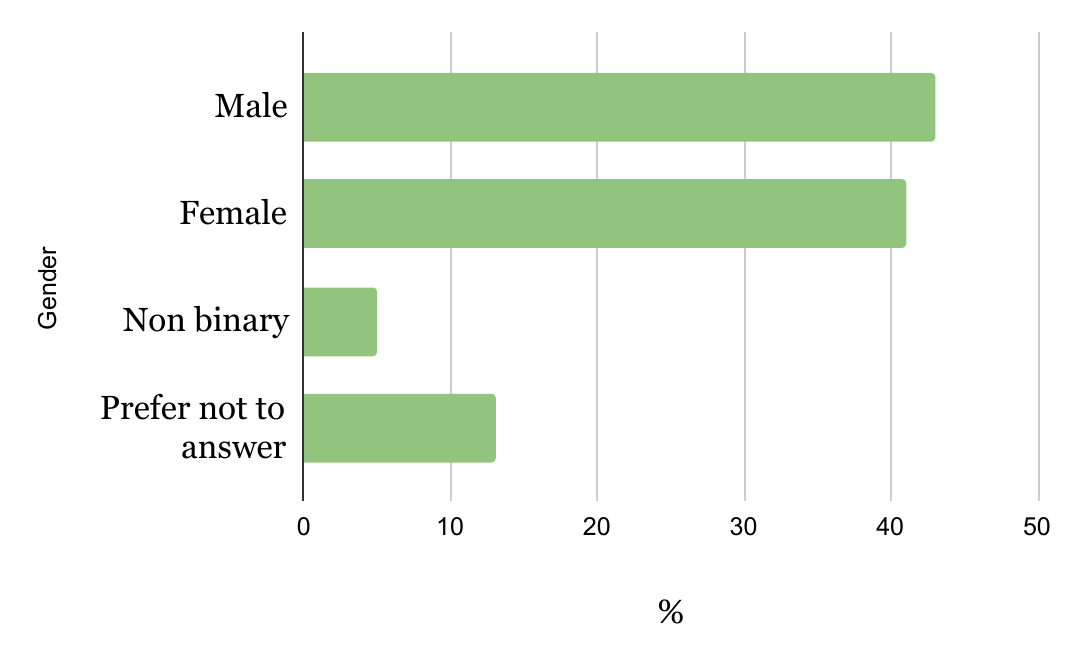}}
\caption{Gender distribution of the participants}
\label{fig:gender}
\end{figure}

We received a total of 102 responses, with the gender distribution depicted in \cref{fig:gender}.

\begin{figure}[!htb]
\centerline{\includegraphics[width=\columnwidth] {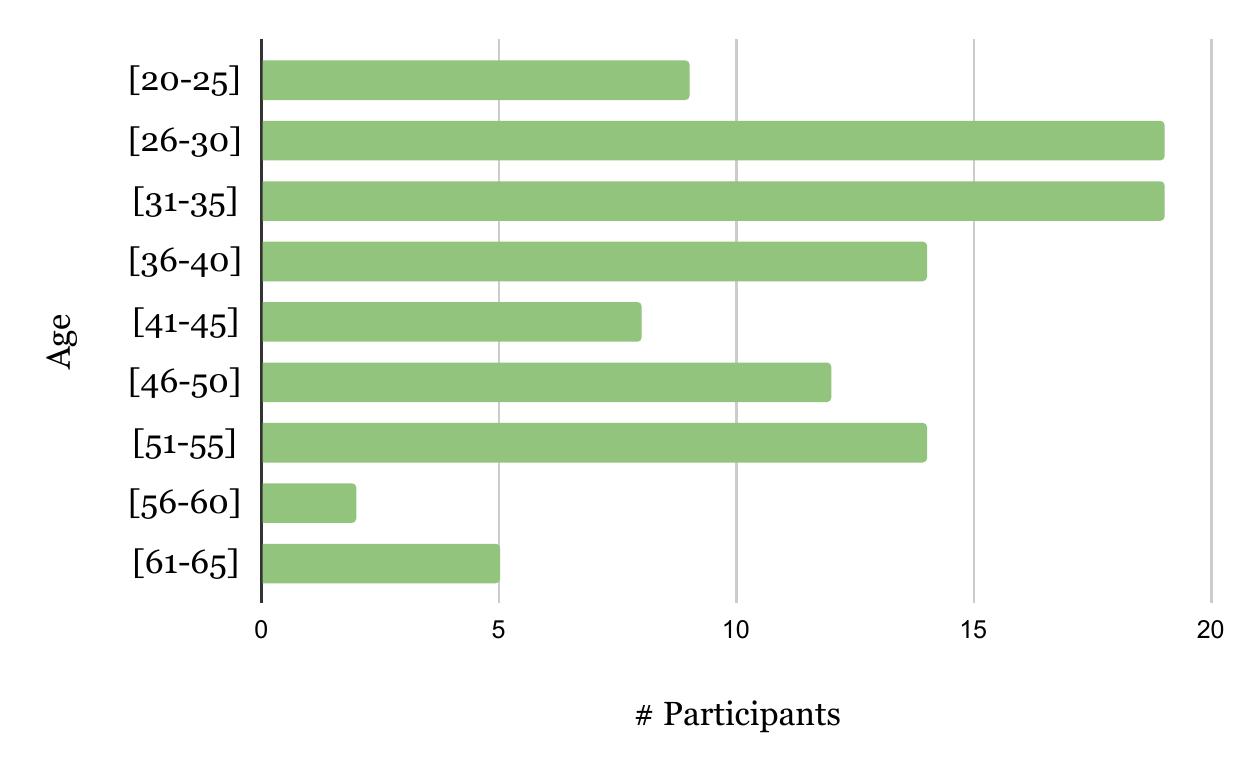}}
\caption{Age distribution of participants (grouped into five-year intervals)}
\label{fig:age}
\end{figure}

\cref{fig:age} presents the age distributions of the participants, grouped into five-year intervals. Interestingly, there are participants in many age ranges, including some older, probably more experienced researchers.

\begin{figure}[!htb]
\centerline{\includegraphics[width=\columnwidth] {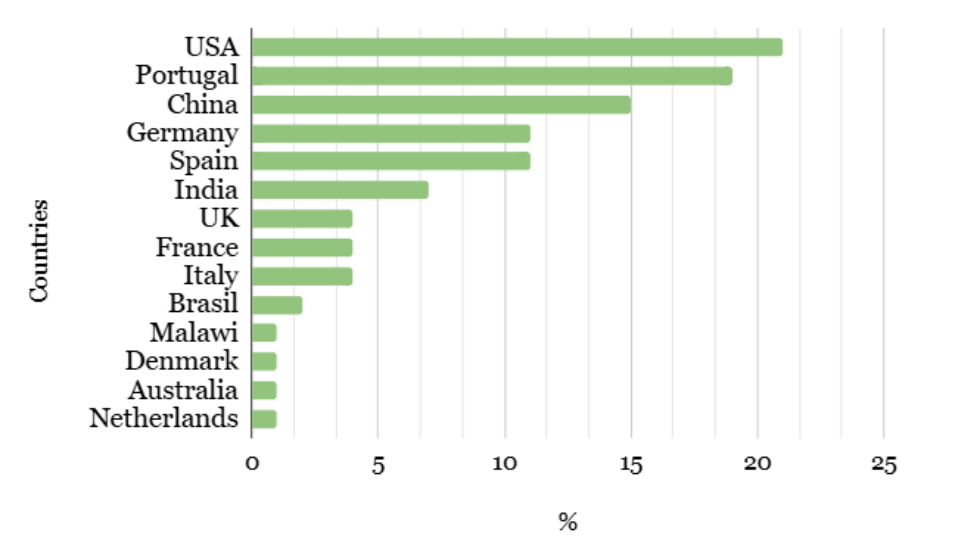}}
\caption{Geographical distribution of participants}
\label{fig:geogra}
\end{figure}

The geographical distribution of the participants is shown in \cref{fig:geogra}. There is a widespread distribution across the globe, with multiple continents represented by many researchers.

\begin{figure}[!htb]
\centerline{\includegraphics[width=\columnwidth] {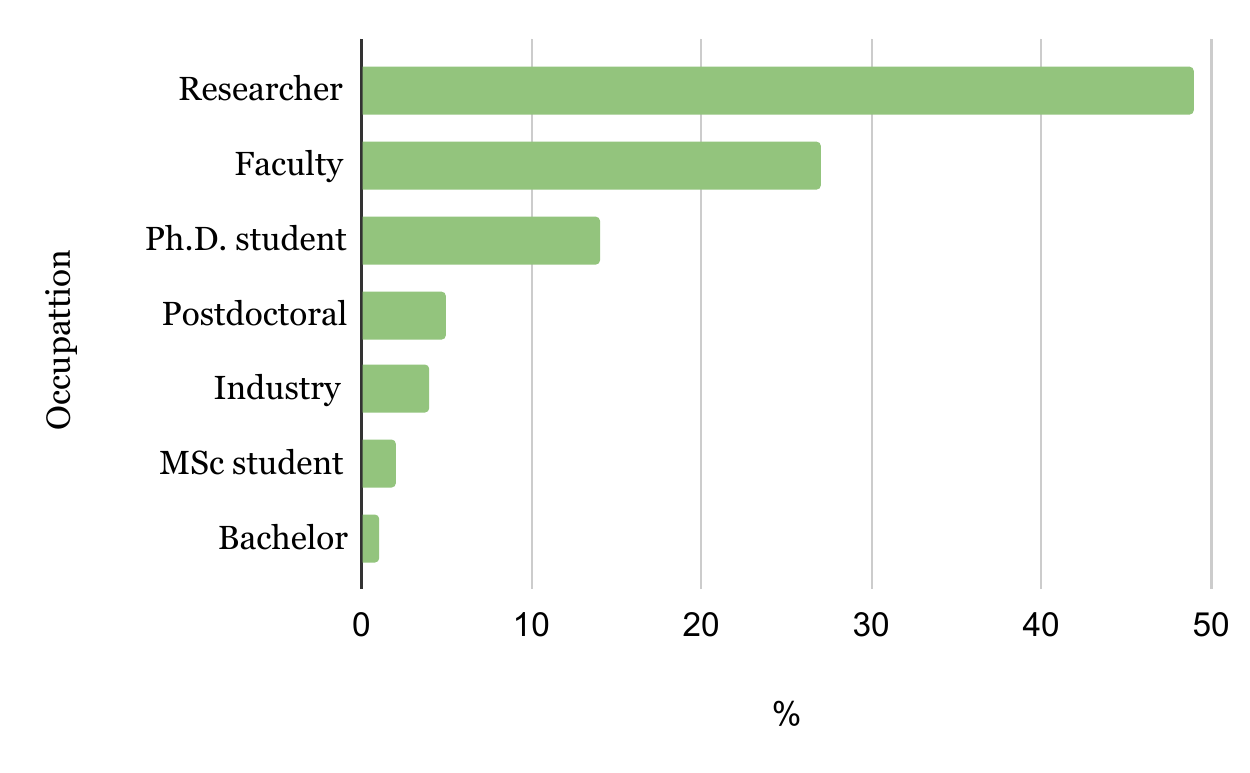}}
\caption{Occupational distribution of participants}
\label{fig:occupation}
\end{figure}

\cref{fig:occupation} illustrates the occupational distribution of the participants. Most participants are researchers or faculty members.

\begin{figure}[!htb]
\centerline{\includegraphics[width=\columnwidth] {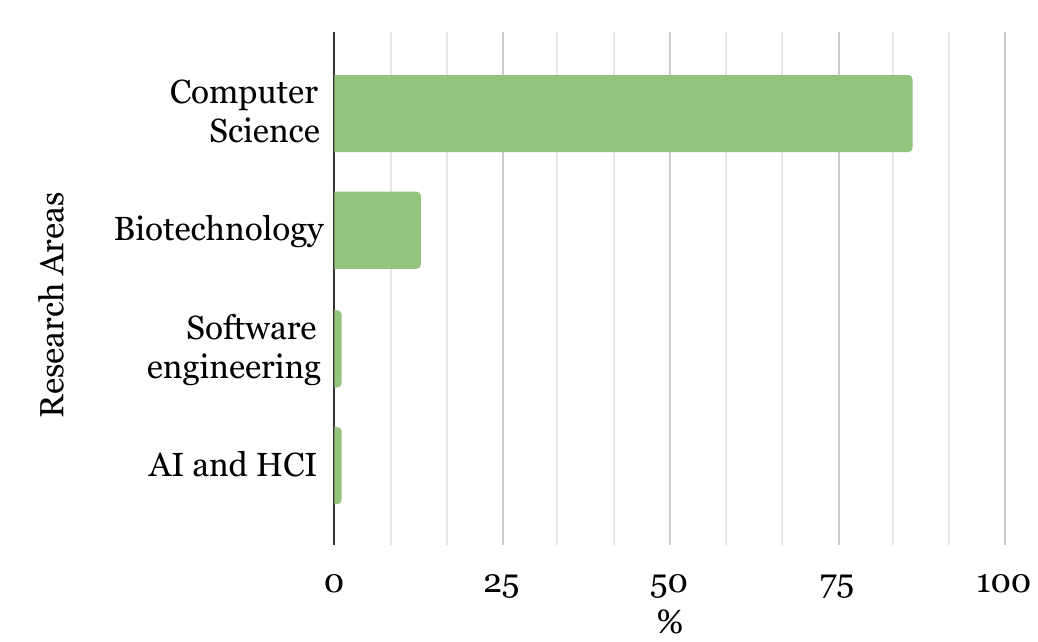}}
\caption{Research field distribution of participants}
\label{fig:areas}
\end{figure}

The distribution of participants across research fields is depicted in \cref{fig:areas}. Not surprisingly, most participants identified themselves as working in computer science, but some made more specific choices. Some participants were from other fields, such as biotechnology.

\begin{figure}[!htb]
\centerline{\includegraphics[width=\columnwidth] {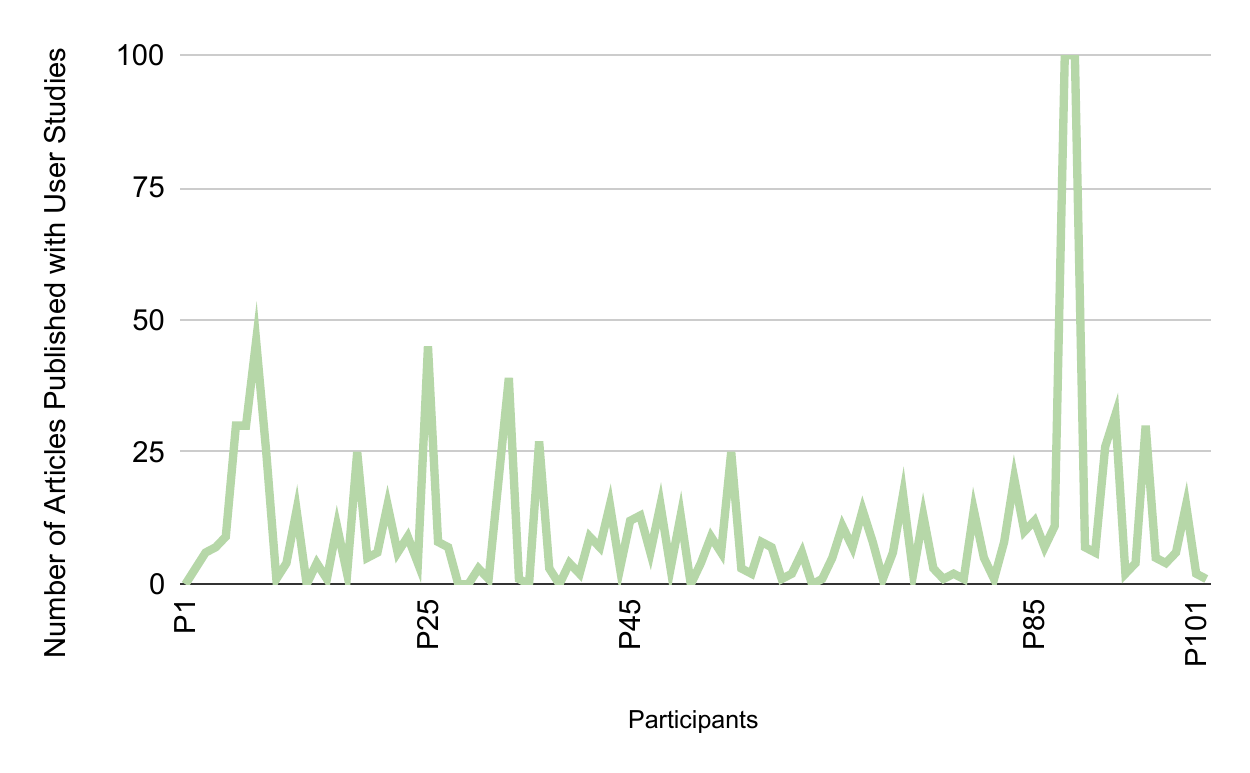}}
\caption{Number of articles published based on user study participant responses}
\label{fig:co-authored}
\end{figure}

In \cref{fig:co-authored}, we present a comparison at the individual level of the number of user study articles that were published. The count of articles incorporating user studies varied from 0 to 100, with a median of 6.

\subsubsection*{Participants' Answers}

\begin{figure}[!htb]
\centerline{\includegraphics[width=\columnwidth] {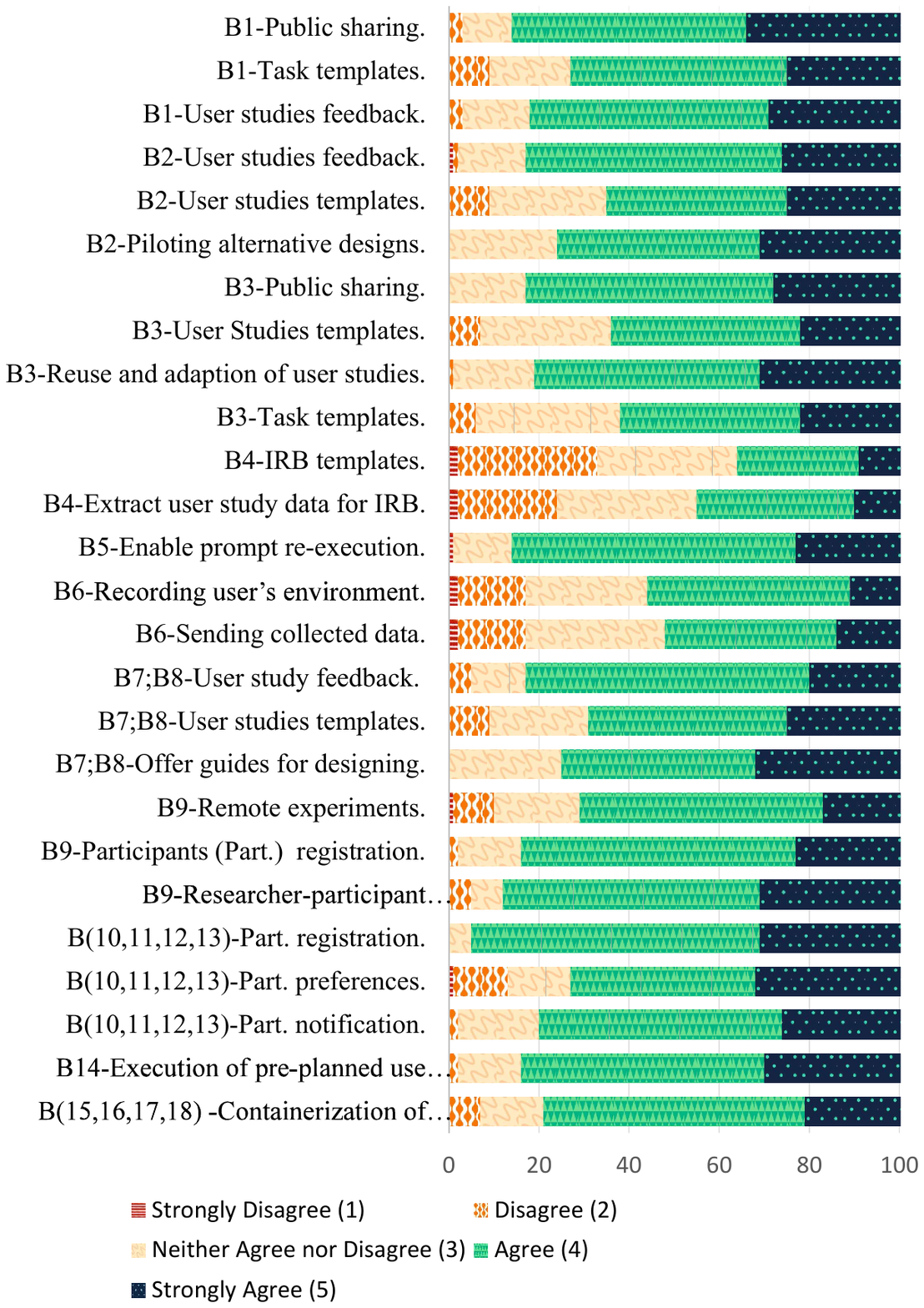}}
    \caption{Distribution of participants' answers}
\label{fig:quartile}
\end{figure}

In \cref{fig:quartile}, we present the participants' answers for each barrier/feature. 
The results show that all features/barriers, except ``$B_4$-\featureIDWithoutDot{IRBTemplates-Our}'' and ``$B_4$-\featureIDWithoutDot{AutomaticIRBInformation-Our}'', have more than 50\% answers in which participants assigned values of 4 or 5 (agree or strongly agree), and the median has a value of 4 (agree). For the other two features/barriers, the median value was 3 (neutral). 
Moreover, for 16 out of the 30 features, more than 75\% of the answers were positive (agree or strongly agree). A negative response is defined as strongly disagree or disagree, and a neutral response is defined as neutral.
Since the median is not influenced by extreme values or outliers in the data, the central or typical value of the data is positive.


\subsubsection{Hypotheses Testing}\label{sec:Userhyptest}
To test our hypotheses, we group and count the numbers of ``Strongly Disagree (1)'', ``Disagree (2)'' and ``Neither Agree nor Disagree (3)'' in a group designated negative/neutral (Neg/Neu)  
and ``Agree (4)'' and ``Strongly Agree (5)'' in another group, designated the positive (Pos) group.

To formally evaluate our hypotheses, we define the following family of null hypotheses, where $f$ is each of the features:

\begin{flushleft}
$H_{0_f}$: The frequency of the Neg/Neu answers is the same as that of Pos answers.
\end{flushleft}

These hypotheses aim to assess whether participants perceive each feature 
as equally effective (positive) or hold neutral or negative views about its ability to help researchers overcome the associated barriers.

We use the chi-square test with a significance level of 0.05 (5\%) to quantitatively evaluate the relationship between the number of responses from the two groups. Specifically, we test whether the proportion of positive responses for each feature is significantly different from the combined proportion of negative and neutral responses.

The chi-square test can compare frequency distributions between different groups or categories and can be applied to ordinal data, such as that from a Likert scale.
The chi-square test is robust enough to be applied even when the data do not follow a normal distribution, which is the case for our data~\cite{peters1940chi}.

In \cref{tab:chiSquare}, we show the number of responses for each feature in each group and the p-value for the chi-square test.

\begin{table*}[!tbh]
  \centering
  \caption{Distribution of answers across the two predefined groups (1, 2, 3 and 4, 5) and chi-square test results}
    \begin{tabular}{m{7.55cm}>{\raggedright}m{5.05cm}>{\centering}m{1.2cm}>{\centering}m{.8cm}>{\raggedright\arraybackslash}m{.65cm}}
    \toprule
    
    \textbf{Barrier} & \textbf{Feature} & \textbf{Neg/Neu Score 1-3}& \textbf{Pos Score 4-5} & \textbf{p-value}\\
    
    \midrule
    \multirow{3}{*}{$B_{1}$: Task design is hard.} &     \featureID{public} & 14 & 88 & 0 \\
    & \featureID{taskTemplates} &27 & 75 & 0 \\
     & \featureID{userStudiesFeedback-Our} &18 & 84 & 0 \\

    \midrule
    \multirow{3}{*}{\parbox{7.55cm} {$B_{2}$: Difficult to understand design trade-offs in advance.}}  & \featureID{userStudiesFeedback-Our} &17 & 85 & 0 \\
    & \featureID{userStudiesTemplates} & 36 & 66 & 0.002 \\
     & \featureID{piloting-Our} & 24 & 78 & 0 \\

    \midrule
    \multirow{4}{*}{\parbox{7.55cm} {$B_{3}$: Can not take any study off the shelf.}}  &     \featureID{public} & 17 & 85 & 0 \\
     & \featureID{userStudiesTemplates} & 36 & 66 & 0.003 \\
     & \featureID{reuseAdapt} & 19 & 83 & 0 \\
     & \featureID{taskTemplates} & 38 & 64 &0.01 \\

    \midrule
    \multirow{2}{*}{\parbox{7.55cm} {$B_{4}$: Institutional review board (IRB) requires unnecessary effort.}}  & \featureIDBold{IRBTemplates-Our} & \textbf{64} & \textbf{38} & \textbf{0.01} \\
      & \textit{\featureID{AutomaticIRBInformation-Our}} & \textit{55} & \textit{47} & \textit{0.43} \\

    \midrule
    $B_{5}$: Building/integrating tools is challenging. & \featureID{promptReExecution-Our} & 14 & 88 & 0 \\

    \midrule
    \multirow{2}{*}{\parbox{7.55cm} {$B_{6}$: Difficult to get data collection right.}} & \textit{\featureID{recordEnvironment}} & \textit{44} & \textit{58} & \textit{0.17} \\
     & \textit{\featureID{sendCollectedData}} & \textit{48} & \textit{54} & \textit{0.55} \\

    \midrule
    \multirow{3}{*}{\parbox{7.55cm} {$B_{7}$: Lack of knowledge (empirical evaluation).\\ $B_{8}$: Gaining the needed knowledge is inefficient.}}
     &  \featureID{userStudiesFeedback-Our}  & 17 & 85 & 0 \\
     & \featureID{userStudiesTemplates}  & 31 & 71 & 0 \\
     & \featureID{offerGuides-Our} &     26 & 76 & 0 \\

    \midrule 
    \multirow{4}{*}{\parbox{7.55cm} {$B_{9}$: Researchers are uncomfortable with people.}} & \featureID{remoteExperiments} & 29 & 73 & 0 \\
     & \featureID{register} &16 & 86 & 0 \\
     & \featureID{communication} &  12 & 90 & 0 \\
          
    \midrule 
     \multirow{4}{*}{\parbox{7.55cm} {$B_{10}$: Hard to recruit participants.\\
     $B_{11}$: Hard to manage participants.\\
$B_{12}$: Recruiting norms vary by study/org.\\
 $B_{13}$: Hard to select participants.}} 
     & \featureID{register} & 5 & 97 & 0 \\
     & \featureID{participantsPreferences} & 27 & 75 & 0 \\
     & \featureID{notifyParticipants} & 20 & 82 & 0 \\
     &  &  & &  \\

    \midrule
    $B_{14}$: End-to-end orchestration is cumbersome. & \featureID{PrePlanesUserStudy-Our} & 16 & 86 & 0 \\
    
    \midrule
    \multirow{4}{*}{\parbox{7.55cm}{
    $B_{15}$: Prototype software is not ready for deployment.\\ 
    $B_{16}$: Deploying to a participant’s local personal computer is challenging.\\ 
    $B_{17}$: Deploying to hosted virtual machines is challenging.\\  $B_{18}$: Deploying to the web is challenging.}} & \featureID{containerization-Our} &  21 & 81 & 0 \\
    &  &  & &  \\
    &  &  & &  \\
    &  &  & &  \\
    &  &  & &  \\

    \bottomrule
        \end{tabular}%
  \label{tab:chiSquare}%
\end{table*}%

The p-value is lower than 0.05 for all but three features (highlighted in italics in the table): ``\featureIDWithoutDot{AutomaticIRBInformation-Our}'', ``\featureIDWithoutDot{recordEnvironment}'', and ``\featureIDWithoutDot{sendCollectedData}'', corresponding to barriers $B_4$ and $B_6$. Thus, we can reject the null hypothesis $H_{0_f}$ for all the other features.

For these three features, we proceeded with further analysis. We divided the answers into 3 groups: \textit{Neg} ``Strongly Disagree (1)'' and ``Disagree (2)'',  \textit{Neu)} ``Neither Agree nor Disagree (3)'',  \textit{Pos)} ``Agree (4)'' and ``Strongly Agree (5)''.
Moreover, for these three features, we define the following new null hypotheses:

\begin{flushleft}
$H'_{0_f}$: The frequency of the Neg answers is the same as that of Pos answers.
\end{flushleft}



In \cref{tab:chiSquare3Groups}, we present the number of responses for these three features for each of the three answer groups (Neg, Neu, Pos), as well as the p-value for the chi-square test. 
In fact, we also include the feature ``\featureIDWithoutDot{IRBTemplates-Our}'' because, although there is a statistically significant difference between Neg/Neu and Pos, we want to further explore the relationship between strictly negative and positive answers.
The column labeled ``p-value Neg/Pos'' represents the significance level of the difference between the negative and positive groups. 

\begin{table*}[!htb]
  \centering
  \caption{Distribution of answers across the three predefined groups for the three features and chi-square test results pairwise (positive/negative)}
    \begin{tabular}{m{5.3cm}>{\raggedright}m{4.4cm}>{\centering}m{0.7cm}>{\centering}m{0.7cm}>{\centering}m{0.7cm}>{\centering\arraybackslash}m{1.3cm}}
    \toprule
\textbf{Barrier} & \textbf{Feature} & \textbf{Neg Score 1-2}&\textbf{Neu Score 3} & \textbf{Pos Score 4-5} & \textbf{p-value Neg/Pos} \\
    \midrule
     \multirow{2}{*}{\parbox{5.3cm} {$B_{4}$: Institutional review board (IRB) requires unnecessary effort.}} & \featureID{IRBTemplates-Our} & {33}    & \textit{31}    & {38}    & {0.48} \\
     & \featureID{AutomaticIRBInformation-Our} & 24    & 31    & 47    & 0.005 \\
    \midrule
	\multirow{2}{*}{\parbox{5.3cm} {$B_{6}$: Difficult to get data collection right.}} & \featureID{recordEnvironment} & 17    & 27    & 58    & 0  \\
		& \featureID{sendCollectedData} & 17    & 31    & 54    & 0 \\
    \bottomrule
    \end{tabular}%
  \label{tab:chiSquare3Groups}%
\end{table*}%


On the basis of these results, we reject the null hypothesis $H'_{0_f}$ for three features, meaning that there is indeed a difference between the positive and negative answers from the survey participants.
However, we cannot reject the null hypothesis for the feature ``\featureIDWithoutDot{IRBTemplates-Our}''.

\subsection{Analysis} \label{subsection:features-analysis}

\subsubsection{Quantitative Analysis}
The null hypotheses we defined (one for each feature) can be discarded 
for all except one feature. 
This means that there is indeed a statistically significant difference between the number of positive and
negative answers. This is even the case when considering a negative answer the neutral answer, except for three features. According to \cref{fig:quartile}, researchers support the features we propose.
This is, however, not the case for one feature, the templates for IRBs.
Although there was a statistically significant difference between the Neg/Neu group and the Pos group (\cref{tab:chiSquare}), considering the data in \cref{fig:quartile}, we can see that the Neg/Neu group was favored only slightly. However, when the neutral group was removed, that is, when only the Neg and Pos groups were considered, we observed no statistically significant difference (\cref{tab:chiSquare3Groups}).

These results suggest that a significant, very experienced part of the research community agrees with the features we propose to address the current barriers.

This is a relevant result, as it can be seen as a list of actionable requirements that can be immediately implemented in the current or future tools to aid researchers in designing and executing user studies.

\subsubsection{Qualitative Analysis}
Survey responses ranged from strong agreement to skepticism regarding certain features. Features like ``\featureIDWithoutDot{userStudiesFeedback-Our}'' and ``\featureIDWithoutDot{public}'' received broad support for fostering collaboration and iterative study design refinement. In contrast, features such as ``\featureIDWithoutDot{IRBTemplates-Our}'' and ``\featureIDWithoutDot{AutomaticIRBInformation-Our}'' elicited mixed reactions due to institutional variability and the complexities of formal review processes. 

Several factors likely contributed to the differing levels of acceptance, including participants' demographic backgrounds, which may influence their perceptions. One participant from the USA provided the following insight:

\begin{flushleft}
\it
``IRB policies and forms seem to vary widely across institutions and even (in our own case) year to year and person to person.''
It seems like this would be hard to automate. Furthermore, the IRB is typically consulted prior to doing anything since the rules seem to change all the time. So [feature] F8 does not seem feasible given the IRB is often incorporated into the process early for these reasons.''
\end{flushleft}

This feedback underscores the complexity and variability of IRB procedures, emphasizing the challenges of automating related processes.

Similarly, another participant expressed skepticism regarding the feasibility of automatic extraction:

\begin{flushleft}
\it
``F8 automatic extraction would be helpful, but I'd be skeptical of how reliable it would be.''
\end{flushleft}

These comments highlight the challenges of creating solutions to streamline IRB processes, as researchers often encounter variability and shifting requirements. Clearly, further work is necessary to address this barrier effectively.

Another significant challenge raised by researchers involves receiving feedback from peers, especially during the early stages of platform adoption. As one participant noted:

\begin{flushleft}
\it
``It would be really hard to get (F3) feedback from others [researchers], unless it is already an active site.''
\end{flushleft}

This underlines the importance of fostering an active community of users to make feedback mechanisms viable and effective.

Some researchers are also skeptical about the applicability of pre-existing user study templates to their specific research needs. For instance, one researcher explained:

\begin{flushleft}
\it
``For most of my studies, I am a bit skeptical that I could leverage a study that someone else has designed as a starting point, as so many of our studies have required careful tailoring to the tool and the research questions we have.''
\end{flushleft}

This reflects the broader challenge of designing adaptable templates that address a diverse range of research needs. Nevertheless, others see value in this kind of approach. For instance, one researcher emphasized the value of centralizing existing studies to improve accessibility and understanding:

\begin{flushleft}
\it
``Somewhat tangential, it feels like another value of a platform would be a sort of centralization. There are so many different studies of developers across disciplines, it's hard to know all of what's out there and what the results were.''
\end{flushleft}

Interpreting data has also emerged as a key challenge. One participant highlighted the difficulty of deriving meaningful insights from user behavior:

\begin{flushleft}
\it
``One challenge is how to interpret the data. Making sense out of what the user is doing is often non-trivial and manual, particularly in a non-think-aloud study.''
\end{flushleft}

Despite these challenges, some researchers have downplayed their significance, particularly for barriers such as 
$B_2$, $B_6$, and $B_9$, respectively:

\begin{flushleft}
\it
``Some of these may be more or less appropriate depending on the experience of the researcher. For instance, [feature] F4 may be more valuable to junior researchers than to senior researchers.''
\end{flushleft}

\begin{flushleft}
\it
``F10 and F11 would be nice to have, but Zoom works pretty well. It's also possible to set up a VM with the necessary tools. It's a pain to do, but possible.''
\end{flushleft}

\begin{flushleft}
\it
``This one just doesn't strike me as a major barrier.''
\end{flushleft}

Recruitment is widely regarded as one of the most challenging aspects of user studies. Participants identified problems related to motivating participants and managing incentives:

\begin{flushleft}
\it
``This sounds useful, but suffers from the problem of why devs would sign up for the platform. What benefit do they get? Why would people do repeated developer studies? For my past research on devs, recruiting is one of the biggest problems we faced, and it took tremendous effort.''
\end{flushleft}

\begin{flushleft}
\it
`Consider the platform managing incentives to participants as well.''
\end{flushleft}

On a more positive note, participants appreciated the containerization of user studies, recognizing its potential to simplify experimental setups. Feedback included:

\begin{flushleft}
\it
``That really depends on the object of the study. If users will execute the software, for sure an easy-to-go environment will help the application.''
\end{flushleft}

\begin{flushleft}
\it
``Container or VM would be really useful, since devs have so many different devices.''
\end{flushleft}

Furthermore, some participants were not so eager to adopt features such as ``\featureIDWithoutDot{recordEnvironment}'' and ``\featureIDWithoutDot{sendCollectedData}''. 
For these cases, we did not receive relevant comments, but one participant agreed that Zoom works well, and another agreed that the privacy of data collection might be at stake. Thus, integrating a mechanism such as Zoom into a tool seems to be a possible path.

In summary, while the proposed features address many barriers, specific challenges--such as IRB processes, researcher feedback, recruitment, and study personalization--require further refinement to meet the diverse needs of the research community. Despite some skepticism, participants highlighted promising aspects, such as containerization, which could significantly simplify the execution of user studies. Survey participants did not suggest additional features.

\subsection{Answering RQ2} \label{subsection:features-rq2}
To answer RQ2, we proposed a set of features to address barriers faced by researchers when designing and running user studies. Some of these features were already existing in current tools, while other features were devised by us.
For all the features except one, the number of positive responses is significantly greater than the number of negative responses, indicating that the proposed features are supported by the research community. 


\section{Threats to Validity} 
\label{sec:Userthreats}

In this section, we discuss several potential threats to the validity of our study, which are divided into four sections~\cite{Cook1979, Wohlin2012}, which are addressed next.


\subsection{Internal validity}
The effect of natural variation on human performance is based on how subjects are selected from a larger group, leading to potential biases. The self-selection of researchers to answer the questionnaire can introduce motivation and appropriateness biases, making the selected group less representative  of the entire user studies population. Notably, a significant majority (76\%) of our respondents are researchers with over two years of experience in developing and publishing user studies. Moreover, rather than making the study publicly available on social media, we have sent it directly (through email) to researchers who published at conferences where user studies are common (CHI, ICSE, and VL/HCC). Importantly, no compensation was offered. Thus, we can assume that our participants were interested in providing their knowledge to our study. Thus, it is likely that our population represents experts in the subject, thus making them relevant respondents.

\subsection{Construct validity}
Researchers' beliefs or expectations about the study outcomes can inadvertently influence participant behavior or the interpretation of results. 
However, we have defined quantitative methods as the main 
driver for the analysis and, consequently, for the results.

The interpretation of the barriers and features by the participants may have influenced their answers. For instance, they may not have fully realized the extent of the barrier or the benefit of the feature. However, the results are quite consistent, and no researcher has raised that issue in the open-ended questions.

\subsection{External validity}

The sampling bias may also influence the external validity of our study. Our survey respondents were recruited from a specific pool of researchers and practitioners in the fields of SE and human-centered computing. This may limit the generalizability of our findings to a broader population of researchers outside these domains.
We attempted to mitigate this threat by leveraging diverse recruitment strategies, including outreach through social media and different research conferences, to capture a more representative sample of the target population.

Our study focused on a selection of platforms relevant to prior research on user studies for Software Engineering~\cite{Myers2023}. While other tools exist~\cite{Heitz2024, Daniel2023, Schwind2023, Costa2020}, we deliberately limited our scope to ensure a focused and feasible feature analysis. Future work could expand this study by systematically reviewing a broader range of tools across different domains to identify additional relevant features.

\subsection{Conclusion validity}

Our statistical analyses rely on certain assumptions, such as the independence of observations and the appropriateness of the chi-square test for categorical data. Violations of these assumptions could affect the validity of our statistical conclusions. However, we strive to prevent such violations and have determined that the chi-square test is the most appropriate for our data type.


While our study confirms that researchers generally perceive the proposed features as useful, we do not claim that they fully eliminate the identified barriers. User studies are inherently complex, and practical challenges often require multiple complementary solutions. In our survey, we explicitly asked participants: ``Please add below any comments you might have on the proposed features and/or suggest another feature to overcome the previous barrier.'' No additional features were suggested, which may indicate that the proposed features align well with the researcher's needs. However, this does not necessarily mean that participants viewed them as fully resolving all barriers. Instead, the features should be understood as practical steps toward mitigating challenges rather than absolute solutions.

Another potential threat to validity is that participants may have interpreted the survey questions differently based on their assumptions about feature completeness. The survey did not explicitly state whether participants should assume that all features would be fully implemented at launch. For instance, some may have assumed that features like task templates required a comprehensive, predefined library, while others may have considered a more dynamic, evolving system where researchers could contribute templates over time. This variation in interpretation could have influenced responses, leading to differences in perceived feasibility and impact.

\section{Discussion} 
\label{sec:Userdiscussion}



In this section, we discuss our findings and the feasibility of implementing the proposed features, the challenges associated with different levels of complexity, and the long-term sustainability of a tool that implements these features.

During the analysis of the ten existing tools to aid researchers in designing and running user studies, we identified a total of 15 different features. However, the characterization of tools by features involves a certain subjectivity, which can lead to a different characterization when carried out by another researcher. Nevertheless, the new features we propose are not available in any of the tools, and except for one, all were considered relevant by the research community in the survey we conducted.

While the proposed features were generally well received, their implementation varies in complexity. Some, such as offering guides for study design and enabling researcher feedback, are relatively straightforward, requiring structured content and researcher collaboration. These could be introduced early in development, allowing the research community to refine resources dynamically. More technically demanding, are features like the automatic execution of pre-planned studies and prompt re-execution, which necessitate a sophisticated infrastructure for scheduling, execution, and monitoring. Implementing these capabilities would require integration with experiment orchestration tools\cite{Ludascher2009, Freire2012, Hanken2015Yesworkflow} or workflow automation systems~\cite{Schaefer2013, Wettinger2016} to support reproducible studies across different environments.

Among the most promising yet technically challenging features is containerization, which simplifies deployment and ensures reproducibility across diverse computing systems. While this approach reduces the burden of manual experiment setup, fully automating deployment remains difficult. Rather than eliminating deployment challenges entirely, the proposed features aim to ease the process. Our prior work~\cite{costa2025backend} explores methods to reduce deployment complexity through containerization and automated setup mechanisms, making it more manageable compared to entirely manual approaches. However, careful planning is necessary to ensure containerized environments remain adaptable to a wide range of research needs while maintaining their usability.

Extracting user study data for IRB submissions presents another challenge due to institutional variability, making full automation impractical. A semi-automated approach is more feasible and could significantly reduce the researcher's workload. Implementing predefined templates for structured report generation is relatively straightforward, as it primarily involves a well-designed form-based system that formats key study details automatically. 
Ensuring compliance through validation mechanisms adds another layer of complexity, as the system would need to accommodate evolving IRB policies across institutions. The most difficult aspect would be integrating such a tool with existing research management platforms, as it would require interoperability with various data sources and the ability to adapt to changing compliance requirements. While complete automation remains unlikely, a flexible, researcher-driven solution that provides structured guidance and customizable templates would be both feasible and highly beneficial. This could be further enhanced through the use of modular plugins tailored to specific institutional requirements, such as those seen in behavioral research platforms like jsPsych~\cite{Leeuw2015}.

The feasibility of these features also depends on whether the platform is designed as an open, researcher-driven system. A practical and sustainable strategy would be an incremental approach, starting with a core framework that allows researchers to contribute and refine study templates over time. Rather than launching with a fully comprehensive feature set, a system that evolves based on community input would ensure adaptability and long-term relevance. Features reliant on researcher engagement, such as study feedback and piloting support, could be introduced early and refined iteratively based on participation.

Developing such a tool also requires addressing long-term maintenance challenges, particularly for features dependent on external services and APIs. Automated data extraction, for example, would require continuous updates to maintain compatibility with evolving metadata standards and publication formats. Similarly, integrating third-party recruitment platforms or cloud-based execution environments would necessitate ongoing modifications to accommodate API changes and authentication protocols. Ensuring the platform’s sustainability will require a modular and extensible design, allowing for incremental updates and researcher-driven enhancements over time.

In summary, the features we proposed and collected to address these barriers were mostly well accepted by the 102 participants of our survey. We could also verify that no current tool can cope with all the features, not even those existing in other tools. Thus, with our work, we define a list of actionable requirements for current tools to evolve and better support researchers when working on user studies. Moreover, these can also serve as guides for future tools.

\section{Conclusion} 
\label{sec:Userconclusion}
On the basis of the literature analysis and current tools, we propose a list of key tool features that can support the execution of all the stages of user studies.
However, not all the barriers faced by researchers reported by \citet{Myers2023}
are addressed by existing tools. Thus, it is necessary to design, implement, and maintain new tools that can properly support researchers.

In future work, we intend to implement our own tool on the basis of this work. Moreover, we will extend this analysis to the challenges faced by researchers when preparing reproducibility packages \cite{Ivie2018} and the existing tools to address such problems. 
\balance



\printcredits

\bibliographystyle{cas-model2-names}

\bibliography{bibliography}

\end{document}